\documentclass[aip,twocolumn,groupedaddress]{revtex4}
\usepackage{graphicx}
\usepackage{bm}
\usepackage{times}
\usepackage{amssymb}
\usepackage{amsmath}

\begin{document} 

\title{Diffraction imaging for in-situ characterization of double-crystal x-ray monochromators\footnote{submitted for publication to Journal of Applied Crystallography http://journals.iucr.org/j/}}

\author{Stanislav Stoupin}
\email{sstoupin@aps.anl.gov}
\affiliation{Advanced Photon Source, Argonne National Laboratory, Illinois, USA} 
\author{Zunping Liu}
\affiliation{Advanced Photon Source, Argonne National Laboratory, Illinois, USA} 
\author{Steve M. Heald}
\affiliation{Advanced Photon Source, Argonne National Laboratory, Illinois, USA} 
\author{Dale Brewe} 
\affiliation{Advanced Photon Source, Argonne National Laboratory, Illinois, USA} 
\author{Mati Meron}
\affiliation{CARS, The University of Chicago, Illinois, USA}

\date{\today} 

\begin{abstract} 
Imaging of the Bragg reflected x-ray beam is proposed and validated as an in-situ method for characterization of performance of double-crystal monochromators under the heat load of intense synchrotron radiation. A sequence of images is collected at different angular positions on the reflectivity curve of the second crystal and analyzed. The method provides rapid evaluation of the wavefront of the exit beam, which relates to local misorientation of the crystal planes along the beam footprint on the thermally distorted first crystal. The measured misorientation can be directly compared to results of finite element analysis. The imaging method offers an additional insight on the local intrinsic crystal quality over the footprint of the incident x-ray beam.
\end{abstract} 

\maketitle

\section{Introduction}

Double-crystal x-ray monochromators utilizing nearly perfect crystals of various semiconductor materials (primarily Si and Ge) 
are the primary optical devices for monochromatization of intense synchrotron radiation in the photon energy range of 5-30~keV. 
Such monochromators deliver monochromatized radiation to a wide variety of x-ray experiments where 
relative photon energy bandwidths $\Delta E/E \lesssim 10^{-4}$ are required. 
Besides the energy bandwidth the main performance criteria of the double-crystal monochromators 
are the stability and reproducibility of the photon energy scale and beam position, monochromator throughput and spectral efficiency.
Finally, an important criterion to address is the preservation of wavefront, which in the recent years has 
become increasingly important since coherence properties of x-ray sources are drastically improving. 
In this regard, in-situ characterization and understanding of performance parameters of the double-crystal 
monochromators has become a high priority for the synchrotron community.  

A number of reports has been published over the years on the in-situ characterization of performance of various double-crystal monochromators. A major effort was related to minimization of thermal distortion on the first crystal of double-crystal monochromators for third-generation synchrotron sources (e.g., \cite{Smither90, BermanSRN91, Bilderback00}). Solutions involving cryo-cooling of the monochromator crystals have been found and their performance limits investigated (e.g., \cite{WKLee00,WKLee01,Hoszowska01,LZhang03}). To gain detailed understanding on the performance of high-heat-load double-crystal monochromators advanced characterization methods have been developed which combine finite-element analysis of thermo-mechanical response of the monochromator crystals under the heat load of synchrotron radiation with the dynamical theory of x-ray diffraction for distorted crystals (\cite{Mocella03,RHuang14}). 

Advanced in-situ characterization techniques have been applied focused on the evaluation of angular profile of the radiation beam delivered by the monochromators. The actual methods that have been used included measurements of the monochromator crystal angular reflectivity curves (i.e., rocking curves) for different portions of the exit beam (e.g., \cite{Oversluizen89, BermanNIM91,Chumakov04,LZhang13}), measurements using crystal analyzers \cite{Chumakov04}, and x-ray grating interferometry \cite{Rutishauser13}. 

As commonly inferred, these techniques enable evaluation of the slope error on the first crystal due to the thermally induced distortion, which is a crucial step in the performance diagnostics of the high-heat-load monochromators. 
However, the demonstrated methods have not yet become widely spread in the community due to either a time consuming setup and measurement procedure or limited availability of specialized instrumentation such as crystal analyzers or x-ray interferometers. 

\begin{figure*}
\label{fig:setup}
\includegraphics[scale=1.6]{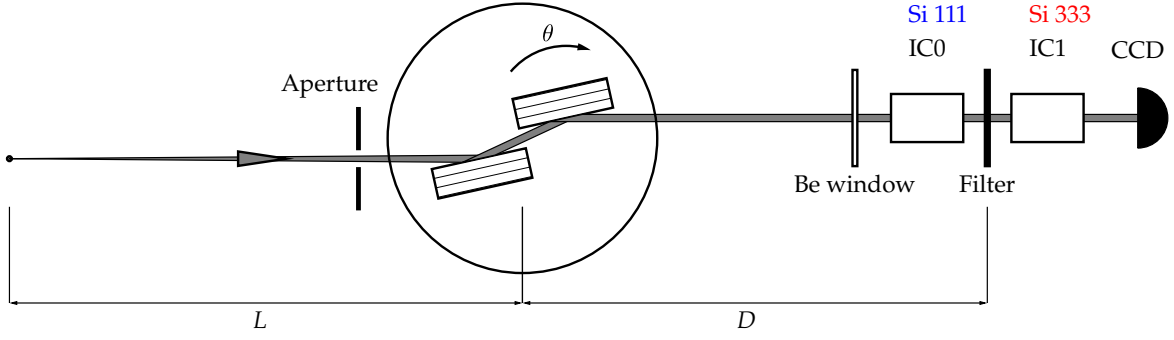} 
\caption{Experimental setup for in-situ performance characterization of double-crystal monochromators (see text for details).} 
\end{figure*}

A commonly used basic method for in-situ performance characterization of double-crystal monochromators involves collection and interpretation of integrated (total) rocking curves at variable levels of power delivered by the x-ray sources. A typical experimental setup is schematically shown in Fig.~\ref{fig:setup}. The intensity of the exit beam dominated by the primary low-order reflection (Si 111) is monitored using an ionization chamber IC0, while radiation reflected by the higher order reflection (e.g., Si 333) is filtered and monitored using an additional ionization chamber IC1. The integrated intensity detected by the ionization chamber is plotted as a function of the Bragg angle of the second crystal which yields Si 111 and Si 333 rocking curves at various levels of power delivered by the x-ray source or power absorbed in the first crystal of the monochromator. 

In this work, an area detector (CCD) is placed after IC1 to produce a sequence of images of the cross section of the exit beam (Si 333) at various angular positions of the second crystal on its rocking curve. It is shown how imaging of the exit beam yields additional insights on the in-situ performance of the monochromator. In particular, it is shown that 
{\it (i)} thermally stabilized crystals in the double-crystal monochromators operate in the regime of weak lattice distortions where ray tracing approach is applicable, which permits unambiguous assignment of the portions of the images to the respective portions of the crystal surface; 
{\it (ii)} at cryogenic temperatures the angular profile of the exit beam in the scattering plane directly relates to the heat-load-induced misorientation of the crystal planes along the x-ray beam footprint; 
and {\it (iii)} sequential imaging of the cross section of the exit beam yields topographs representing local misorientation of the lattice planes and local crystal quality over the incident beam footprint. 
These conclusions are supported by experiments with double-crystal monochromators operated at a bending magnet beamline and an undulator beamline of the Advanced Photon Source. The spatial resolution of the imaging method approaches the x-ray extinction depth in Bragg diffraction from perfect crystals (typically a few microns).
Empowered by diffraction imaging, rapid in-situ evaluation of the wavefront achieved in a single angular scan can facilitate design and implementation of adaptive monochromator systems where the state of the crystal is manipulated to compensate the induced thermal distortions. This can be considered as a viable approach toward wavefront-preserving high-heat-load diffracting optics. 

\section{Theory}\label{sec:theory}
\begin{figure*}
\includegraphics[scale=0.8]{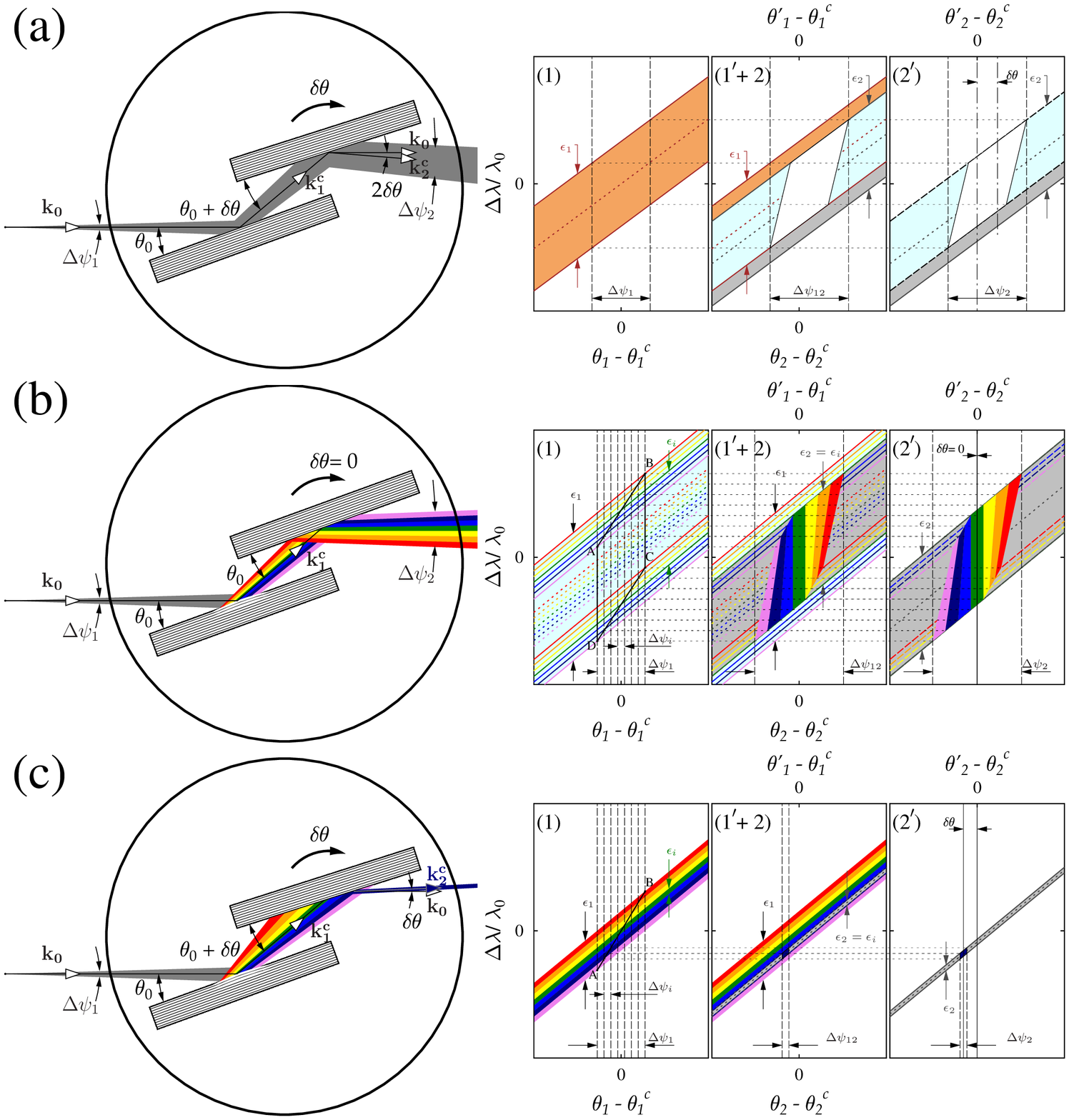} 
\caption{Ray tracing and DuMond diagrams of the double-crystal monochromator (see text for more details):\\
(a) The case of two perfect crystals with an identical symmetric crystallographic orientation. If an angular offset $\delta \theta$ is introduced a partial overlap of the reflection regions of the two crystals takes place in (1'+2). \\
In cases (b) and (c) the distorted surface of the first crystal is considered as a superposition of flat perfect segments with variable angular orientation. X-rays reflected from these segments are schematically shown by variable color. \\
(b) The case of low order reflection (Si 111). The individual segments are presented as a sequence of strongly overlapping reflection regions. \\
(c) The case of high order reflection (Si 333). For simplicity the entrance space of the first crystal is presented as a superposition of non-overlapping local reflection regions. The reflection region of the second crystal selects a portion of radiation which corresponds to a segment with a particular local misorientation depending on the angular offset $\delta \theta$.} 
\label{fig:rtrace}
\end{figure*}

The principle of operation of the monochromator relies on the double-crystal non-dispersive geometry shown in Fig.~\ref{fig:rtrace}(a) where two identical Bragg reflections with equal d-spacing are utilized. 
If the two crystals are perfectly aligned ($\delta \theta$ = 0) the deflection of the incident x-ray beam with a finite divergence $\Delta \psi_1$ due to Bragg reflection from the first crystal is completely compensated by the deflection due to Bragg reflection from the second crystal. The original direction of propagation centered about incident wavevector $\mathbf{k_0}$ is preserved and the divergence of the incident x-ray beam is equal to the divergence of the exit beam ($\Delta \psi_1 = \Delta \psi_2$). 
If an angular detuning $\delta \theta \neq 0$ of the second crystal is introduced as shown in Fig.~\ref{fig:rtrace}(a) a partial overlap of the reflection regions of the two crystals takes place as shown in the corresponding DuMond diagram in the wavelength-angular exit space of the first crystal which is also the entrance wavelength-angular space of the second crystal (panel (1'+2)). The notations on the angular axes of the DuMond diagram is such that 
$\theta_1 - \theta^c_1$ represents the entrance angular coordinate of the first crystal with respect to the center of its reflection region 
$\theta^c_1$. Similarly, $\theta'_1 - \theta^c_1$ is the exit angular coordinate of the first crystal, $\theta_2 - \theta^c_2$ is the entrance angular coordinate of the second crystal and $\theta'_2 - \theta^c_2$ is the exit angular coordinate of the second crystal. Since the second crystal is rotated clockwise $\theta^c_2 = \theta_0 +\delta \theta$. The wavelength axis of the DuMond diagrams is given in the relative units 
$\Delta \lambda/\lambda_0 = (\lambda - \lambda_0)/\lambda_0$.

As a result of the partial overlap, the center of the reflection region in the exit space of the second crystal (panel (2')) shifts by $\delta \theta$ and the central portion of the exit beam (denoted by wavevector $\mathbf{k^c_2}$) propagates at an angular offset $2 \delta \theta$ from $\mathbf{k_0}$ as shown in the ray tracing diagram of Fig.~\ref{fig:rtrace}(a).

The width of the rocking curve of the second crystal results from the convolution of the two identical reflection regions:

\begin{equation}
\Delta \theta_2 = \sqrt{2} \Delta \theta_H,  \ \Delta \theta_H = \varepsilon_{H} \tan{\theta_0}
\label{eq:dth2}
\end{equation}

In this expression $\Delta \theta_H$ is the intrinsic angular width of the Bragg reflection (also known as Darwin width), 
$\varepsilon_{H}$ is the intrinsic relative energy width of the Bragg reflection (a quantity given by the dynamical theory of x-ray diffraction), and 
$\theta_0$ is the central glancing angle of incidence related to the center $\lambda_0$ of the selected wavelength region via the Bragg's law. 

\begin{equation}
\lambda_0 = 2 d_H \sin{\theta_0},
\label{eq:BL}
\end{equation}
where $d_H$ is the d-spacing of the working reflection.

In the practical case when the relative photon energy bandwidth of the x-ray source is greater than $\varepsilon_H$, the relative photon energy bandwidth selected by the monochromator is\footnote{For practical estimation purposes adding these two terms in quadrature is often considered. A more careful evaluation requires integration of expressions given by the dynamical theory of x-ray diffraction.}
\begin{equation}
\frac{|\Delta \lambda|}{\lambda_0} = \frac{\Delta \psi_1}{\tan{\theta_0}} + \varepsilon_H,
\label{eq:dE}
\end{equation}

This relationship can be easily illustrated using the DuMond diagram in panel (1) of Fig.~\ref{fig:rtrace}(a). The total relative bandwidth is the combination of the intrinsic relative bandwidth $\varepsilon_H$ and the divergence of the incident x-rays $\Delta \psi_1$ expressed in relative energy units using the linearized Bragg's law. 

X-ray absorption of the intense incident x-ray beam in the first crystal and dissipation of the absorbed energy leads to a non-uniform temperature distribution and thermal distortion of the crystal (known as the "heat bump") shown schematically in Fig.~\ref{fig:rtrace}(b,c).
In the following we will show that under certain conditions the image of the exit beam cross section is directly related to the thermal distortion of the first crystal along the footprint of the incident beam.

The non-uniform temperature distribution and the resulting thermal distortion lead to a variation in the d-spacing and in the local orientation of the crystal planes involved in the diffraction process.
Differentiation of the Bragg's law yields:
\begin{equation}
\frac{\delta \lambda}{\lambda_0} = \frac{\delta d_H}{d_H} + \frac{\delta \theta}{\tan{\theta_0}} .
\label{eq:dBL}
\end{equation}

At each fixed photon energy a variation in the d-spacing $\delta d_H$ yields an angular deviation 
\begin{equation}
\delta \theta = - \frac{\delta d_H}{d_H} \tan{\theta_0}
\label{eq:dth_dd}
\end{equation}

The influence of such angular deviation on the resulting wavefront distortion can be neglected at cryogenic temperatures in vicinity of zero thermal expansion for the crystal material. This approximation is quite reasonable based on the fact that thermal expansion is rather small at cryogenic temperatures and an assumption that temperatures of the working crystal region do not deviate substantially from the zero thermal expansion temperature.
To illustrate this approximation, we consider a Si crystal of a high-heat-load monochromator under typical operating conditions at a third-generation synchrotron source. We assume that the maximum temperature variation on the first crystal along the beam footprint is on the order of 10~K in vicinity of the Si zero thermal expansion point ($T_0 \approx$~124~K), and that the operating photon energy $E_0 = hc/\lambda_0 \simeq$~5~keV ($\tan{\theta_0} \simeq$~0.4). This brings the crystal close to the upper limit in the operational heat load ($\approx~400$~W of absorbed power delivered by undulator A at the Advanced Photon Source, undulator deflection parameter K~$\approx$~2). Taking into account the temperature dependence of the Si lattice parameter \cite{Reeber75} Eq.~\ref{eq:dth_dd} yields $\delta \theta <$~1~$\mu$rad, which is small compared to the characteristic divergence of the synchrotron undulator source ($\simeq$~10~$\mu$rad).
We note that in a general case the results of measurements in the double-crystal geometry are simultaneously affected by the variations in the d-spacing and the misorientation of the reflecting planes. Thus, the measured angular profile represents a gradient in the wavefront of the outgoing beam. Deconvolution of the two contributions can be accomplished using monochromatization of the incident beam and application of methods of high-resolution x-ray diffraction topography (e.g., \cite{BoTa98}). This however, is not practical for the in-situ performance diagnostics.  

Neglecting the angular deviation due to temperature-induced variation in d-spacing for each photon energy implies that the wavefront is primarily affected by the local misorientation of the Bragg planes. Here, we need to make another important assumption that the crystal is only weakly distorted, such that geometrical optics (i.e., ray-tracing) approach is applicable. Strong crystal distortions lead to generation of new wavefields due to interbranch scattering (e.g, \cite{Authier70,Gronkowski91,Mocella03}). As a result it may become difficult to draw a correspondence between a local region in the cross-section of the exit beam and a local region on the crystal surface. The physical interpretation of the criterion for the smallness of crystal distortion where the ray-tracing approach is applicable is given by Authier \cite{Authier}.
The lattice misorientation over the extinction length should be much smaller than the angular half-width of the intrinsic Bragg reflectivity curve:
\begin{equation}
\frac{\partial (\Delta \theta)}{\partial l} \Lambda_0 \ll \frac{1}{2}\Delta \theta_H .
\label{eq:crit}
\end{equation}
The extinction length is a minimal characteristic length from which x-rays are dynamically reflected. In the Bragg geometry it is given by the real part of the following expression.
\begin{equation}
\Lambda_0 = \frac{\lambda_0 \sqrt{\gamma_0|\gamma_h|}}{|C|\sqrt{\chi_h \chi_{-h}}},
\label{eq:lam0}
\end{equation}
where $\gamma_0$ and $\gamma_h$ are the direction cosines of the incident and reflected waves respectively, $\chi_h$ and $\chi_{-h}$ are the Fourier components of dielectric susceptibility and C is the polarization factor (C=1 for $\sigma$-polarization of the incident radiation).  
The intrinsic angular width $\Delta \theta_H$ as well as the extinction length $\Lambda_0$ are characteristic to a chosen working reflection. 

As frequently reported in the literature (e.g., \cite{WKLee01,LZhang03,Chumakov04,LZhang13}), for the cryogenically cooled Si monochromators the total misorientation of the Bragg planes along the beam footprint on the first crystal is $\Theta \approx$~10~$\mu$rad. The variation in the lattice misorientation in Eq.~\ref{eq:crit} is on the order of $\Theta \sin{\theta_0}/\Delta y$, where $\Delta y \approx$~1~mm is the typical size of the incident x-ray beam in the scattering plane. 
Table~\ref{tab:crit} shows values of the right-hand side of Eq.~\ref{eq:crit} evaluated for the most practical Si 333 reflection at different photon energies. The values of the angular half-width of the reflection $\frac{1}{2}\Delta \theta_{333}$ are also given for comparison. Thus, the criterion of weakly distorted crystal is satisfied for the high-heat-load monochromators under the operating conditions. 

\begin{table} 
\label{tab:crit}
\caption{Numerical values of the right-hand side and the left-hand side of Eq.~\ref{eq:crit} evaluated separately for the Si 333 reflection at different photon energies to illustrate applicability of the weakly distorted crystal approximation. \\
$E_0^{333} = hc/\lambda_0$ - photon energy, \\
$\Lambda_0$ - extinction length, \\
$\frac{1}{2} \Delta \theta_{333}$ - angular half-width. \\
} 
\begin{tabular}{l c c c c c} 
$E_0^{333}$ [keV]                                                  &  15    &  21    &  27    & 33    & 39 \\
$\frac{\partial (\Delta \theta)}{\partial l}\Lambda_0$ [$\mu$rad]  &  0.1   &  0.07  & 0.06   & 0.05  & 0.04 \\
$\frac{1}{2}\Delta \theta_{333}$ [$\mu$rad]                        &  1.7   &  1.2   & 0.9    & 0.7   & 0.6  \\
\end{tabular} 
\end{table} 


Within the framework of the ray-tracing approach the thermally distorted first crystal can be considered as a superposition of flat perfect segments with a variable angular orientation. X-rays reflected from these segments are schematically shown by variable color in the ray-tracing diagrams (Fig. ~\ref{fig:rtrace}(b,c)) and in the corresponding DuMond diagrams as a sequence of overlapping reflection regions. 
The second crystal in the double crystal scheme is assumed to be perfect since it is subjected to a substantially smaller heat load of the x-ray beam Bragg-reflected from the first crystal and scattered radiation. Thus, the reflection region of the second crystal remains unchanged (grey shaded area in DuMond diagrams of Fig.~\ref{fig:rtrace}). Rotation of the second crystal results in the shift of its reflection region across the overlapping regions representing portions of radiation emanating from the first crystal (panels (1'+2) in Fig.~\ref{fig:rtrace}(b,c)).
The resulting interaction is drastically different for the case of low order reflection (e.g., Si 111) shown in Fig.~\ref{fig:rtrace}(b) and the case of higher order reflection (Si 333) as shown in Fig.~\ref{fig:rtrace}(c). These two interactions occur simultaneously at different photon energies and can be observed separately as discussed in the previous section. 

The variation in the misorientation of the first crystal interacting with the incident x-ray beam can be described by a maximum peak-to-valley deviation $\Theta$ from the nominal "zero" value that corresponds to the perfect flat crystal oriented at a glancing angle of incidence $\theta_0$ with respect to the incident beam. 

The intrinsic angular acceptance of the low-order reflection $\Delta \theta^{111} \gtrsim \Theta$. The reflection regions corresponding to different segments of the crystal strongly overlap in the wavelength-angular entrance space of the first crystal as shown in panel (1') of Fig.~\ref{fig:rtrace}(b) (only the boundaries of the individual regions are shown by varying color; the overlap region is shaded by cyan color).
Without loss of generality it is assumed that the misorientation is a monotonic function of a spatial coordinate along the footprint of the incident beam.
The divergence of x-rays reaching each segment of the entrance surface of the first crystal $\Delta \psi_{i}$ is small (within the framework of geometric optics it approaches zero with the size of the segment). The wavelength-angular space selected by the total incident divergence 
$\Delta \psi_1$ can be approximated by a parallelogram ABCD where the points A, B, C, and D are defined as intersections of the angular boundaries limiting the total divergence with the boundaries of the reflection regions of the furthermost segments. In this simplified consideration the total area of ABCD scales as $((\Theta+\Delta \psi_1)/\tan{\theta_0} + \varepsilon_H)$. Thus, a distorted crystal will select a greater photon bandwidth from the incident beam as compared to the bandwidth selected by the perfect crystal (Eq. \ref{eq:dE}). The illuminated total reflection region ABCD is sliced and consequently projected onto the wavelength-angular space (1'+2) for each individual segment. 
For simplicity zero angular displacement of the second crystal from the central angle $\theta_0$ is assumed ($\delta \theta = 0$). Upon subsequent projection onto the exit space of the second crystal (panel (2')) it is seen that the reflected intensities corresponding to the different segments are present simultaneously in the reflected beam and propagate in different (divergent) directions. 
The angular divergence of the exit x-ray beam is given by
\begin{equation}
\Delta \psi_2 = \Delta \psi_1 + \Theta 
\label{eq:dpsi2b}
\end{equation}

This increase in the angular divergence is critical for evaluation of performance of synchrotron beamlines and is often 
interpreted as a virtual source located at a closer distance. In the context of the double-crystal geometry the misorientation of the Bragg planes on the first crystal due to thermal distortion represents slope error on the surface of the first crystal and is the origin of the additional divergence.
The spatial distribution of the slope error variation along the beam footprint can be extracted from the low-order reflectivity if the measurement method has angular selectivity. This is accomplished in practice using crystal analyzers \cite{Chumakov04} or grating interferometry \cite{Rutishauser13}. 

In the case of a high-order reflection (e.g., Si 333) $\Delta \theta^{333} < \Theta$ a qualitatively different picture emerges (Fig.~\ref{fig:rtrace}(c)). For simplicity it is assumed that the entrance space of the first crystal for the higher order reflection can be 
represented as a superposition of non-overlapping local reflection regions. These regions are sequentially illuminated with
the divergent fan of the incident radiation. The line AB shown in panel (1) serves as a guide to the eye for subsequent illumination of the non-overlapping crystal regions. It can be seen that the region ABCD of Fig.~\ref{fig:rtrace}(b) transforms into the line AB under the assumption of zero overlap between the regions and the fraction of the incident divergence $\Delta \psi_{i}$ (corresponding to an individual crystal segment) approaching zero. 
In the exit space of the first crystal (1'+2) the entrance reflection region of the second crystal selects a portion of radiation which corresponds to a certain illuminated region representing a particular local misorientation (or a small range of misorientations) on the first crystal depending on the angular offset $\delta \theta$. The resulting reflection region projected onto the exit space of the second crystal (panel (2')) has an angular width 
\begin{equation}
\Delta \psi_2 = \Delta \theta^{333}
\label{eq:dpsi2c}
\end{equation}

This selected portion of radiation propagates with an angular offset $\delta \theta$ from the original direction $\mathbf{k_0}$ as shown in the ray tracing diagram of Fig.~\ref{fig:rtrace}(c). We note that Eq.~\ref{eq:dpsi2c} is an idealization related to the assumption of non-overlapping discrete regions. In reality, the exit divergence will depend on the relationship of the local slope error to the intrinsic angular acceptance of the high-order reflection $\Delta \theta^{333}$. 

For the purpose of diagnostics using x-ray diffraction imaging the choice of a high order reflection such as Si 333 is particularly convenient since its intrinsic angular width ($\Delta \theta^{333} \simeq 0.5-3$~$\mu$rad) is smaller than the expected total variation in misorientation $\Theta$ while the extinction depth is not particularly large, which ensures that the criterion given by Eq. \ref{eq:crit} is satisfied.

From the analysis given above it is clear that in the absence of plane misorientation all illuminated regions of the first crystal will equally contribute to the image of the exit beam-cross section in the observation plane at any angular offset $\delta \theta$ of the second crystal.
In contrast, if thermal distortion is present regions corresponding to different misoriented segments of the first crystal 
will be highlighted by maximum intensity at different angular positions of the second crystal $\delta \theta$. 
In the case of a monotonic distribution of misorientation as a function of position on the first crystal along the beam footprint the maximum intensity in the resulting image will monotonically shift in the vertical direction as $\delta \theta$ is scanned.\footnote{Similar considerations apply if the slope error is not a monotonic function of the spatial coordinate. The regions corresponding to a unique slope will appear in the resulting image at the corresponding unique angular shift $\delta \theta$. In this case a more complicated picture in terms of beam displacement in the observation plane is expected.}

To draw one-to-one correspondence between a particular location on the crystal and the location in the observation plane the observation plane should be located at a distance $D < \delta x/\Delta \psi_1$ where $\delta x$ is the required spatial resolution. Alternatively, one can scale the image in the observation plane by the factor $L/(L+D)$ to obtain the images of the crystal. It is also noted that the field of view is typically limited by an aperture of size $\Delta y$ placed upstream the monochromator. In order to avoid diffraction on the aperture it is desirable to limit the observation distance $D \ll \Delta y^2/\lambda_0$. For typical aperture sizes $\Delta y \approx$~1~mm and $\lambda_0 \approx 1 \mathrm{\AA}$ aperture-related diffraction effects are expected for D $\approx 10^4$~m.

Quantitative information on the state of the first crystal can be extracted from the sequence of images obtained at different $\delta \theta$ using rocking curve imaging strategy (e.g., \cite{Lubbert00, Stoupin14}). The images are sorted on per-pixel basis and local rocking curves are analyzed for each pixel. Rocking curve topographs can be constructed as maps of local reflected intensity, curve's width and peak position. 
The map of the reflected intensity represents the beam footprint on the first crystal superimposed on the local Bragg reflectivity. The map of peak positions represents gradient of the wavefront in the scattering plane (i.e., the tangential direction)\footnote{The scattering plane is defined by the incident wavevector $\mathbf{k_0}$ and the reciprocal vector of the reflection. The sensitivity of x-ray diffraction to angular deviations from the scattering plane (i.e., in the sagittal direction) is limited}. 
The map of the width provides an insight on crystal quality. For slightly distorted perfect crystals the expected values should be close to the theoretical value $\sqrt{2} \Delta \theta^{333}$. In the following analysis of angular and spatial resolution in the rocking curve topographs is presented. 


\section{Angular sensitivity and spatial resolution}\label{sec:res}
In the double-crystal configuration the second crystal plays the role of a Si 333 analyzer, which selects different portions of the x-ray fan emanating from the first crystal as also noted in earlier studies (e.g., \cite{Chumakov04}).
In principle, the sensitivity of angular measurements using the Si 333 analyzer is not directly related to the intrinsic angular width of the reflection $\Delta \theta^{333}$. It is rather defined by the measurement accuracy of peak positions of the local rocking curves. This accuracy generally depends on the signal statistics and precision/stability of the crystal rotation. In the case of double-crystal monochromator the signal statistics is excellent due to the high photon flux of the exit beam. The angular stability may approach the level of 0.1~$\mu$rad which is about the practical limit for the angular sensitivity.

\begin{figure} 
\caption{Ray tracing diagram in Bragg diffraction illustrating spatial resolution due to finite penetration depth of the wave field into the crystal.} 
\includegraphics[scale=0.8]{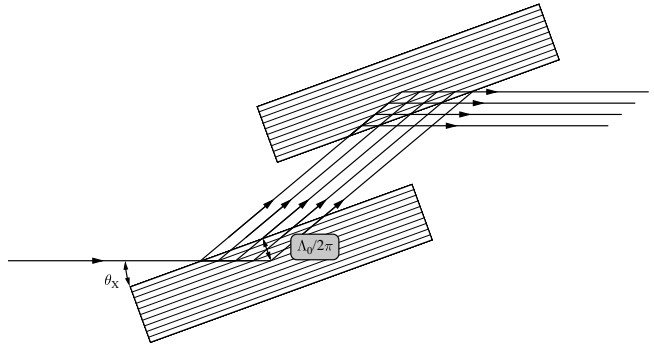}
\label{fig:de}
\end{figure}

The spatial resolution is limited by a number of geometrical factors. 
In practice the diffraction limit can be excluded from the consideration since it is small compared to another fundamental quantity, the wave field penetration depth into the crystal. For an incident wave in Bragg diffraction the penetration depth approaches the extinction depth which attains the minimal value 
$\Lambda_0/2\pi$ in the center of the reflectivity curve. This sets a fundamental limit in the spatial resolution as shown in Fig.~\ref{fig:de}. On the tails of the reflectivity curve the penetration depth into the crystal approaches the absorption depth $\zeta$ (typically $\zeta > \Lambda_0/2\pi$ for practical cases in hard x-ray optics).
Thus, in the measurements of peak position of local rocking curves the spatial resolution in the observation plane due to the finite penetration depth approaches 
\begin{equation}
\delta y_p = \frac{1}{\pi}\Lambda_0 \cos{\theta_0}
\label{eq:y_p}
\end{equation}

For Si, the primary crystal used in high-heat-load monochromators the extinction depth of the 333 reflection is about 4~$\mu$m. 
In the practical range of the primary photon energies 3-15~keV, $\cos{\theta_0} \simeq 1$ and  $\delta y_p \simeq$~ 4 $\mu$m.

Another source of blurring in the images of the exit beam is the finite source size $S$. The related spatial resolution in the observation plane at a distance $D$ from the monochromator is 
\begin{equation}
\delta y_s = D \frac{S}{L}.
\label{eq:y_s}
\end{equation}

For an undulator source at a third generation synchrotron $S \simeq $~25~$\mu$m (FWHM) while for a bending magnet beamline $S \simeq $~70~$\mu$m (FWHM). 

Also, the intrinsic angular deviation of the exit beam, which is present locally for each segment and illustrated for perfect crystals in 
Fig.~\ref{fig:rtrace}(a) has to be taken into account. The related spatial resolution in the observation plane is

\begin{equation}
\delta y_a = D  2\sqrt{2} \Delta \theta^{333}.
\label{eq:y_a}
\end{equation}

The contributions to spatial resolution given by Eqs.~(\ref{eq:y_s}) and (\ref{eq:y_a}) can be optimized by placing the area detector at a sufficiently small distance from the monochromator $D \ll L$. 
The resulting spatial resolution can be estimated as 
\begin{equation}
\delta y = \sqrt{\delta y_p^2 + \delta y_s^2 + \delta y_a^2} .
\label{eq:dy}
\end{equation}

Finally, it is noted that diffraction from single crystals leads to spatial filtering of the object wave~\cite{Davis96}.
Precise interpretation of crystal imperfections in diffraction imaging generally requires solutions of Takagi-Taupin equations \cite{Takagi62,Taupin67}. 

\section{Experimental}
Sequential diffraction imaging of the exit beam of the high-heat-load double-crystal Si 111 monochromators was performed at the Advanced Photon Source using experimental setup shown in  in Fig.~\ref{fig:setup}. The exit beam was filtered to select Si 333 reflection for imaging.  
In the first experiment a water-cooled monochromator operating at a bending magnet 1-BM beamline was studied.
The working Si 111 reflection of the monochromator was set to select photon energies $E_0^{111}$~=6~keV from the incident broadband spectrum of bending magnet radiation. The cross section of the incident beam was limited with 1x1~mm$^2$ aperture located upstream of the monochromator at a distance of 23~m from the source.Stabilization was ensured by repetitive measurements of the Si 333 rocking curve using an integrating detector until narrow curves were obtained with FWHM being close-to the theoretical value of 4.2~$\mu$rad. 

In the second experiment a cryo-cooled (indirect liquid nitrogen cooling) double-crystal monochromator at 20-ID undulator beamline was studied. The working Si 111 reflection of the monochromator was set to a photon energy $E_0^{111}$~=11~keV to select the primary harmonic of the undulator that was tuned accordingly (undulator type A at the Advanced Photon Source). 

In both experiments prior to imaging sufficient time was allowed for thermal stabilization under the heat load of the incident x-rays. A sequence of images of the Si 333 reflection was collected corresponding to different angular positions of the second crystal on its rocking curve. 
The temperatures of the crystals were monitored using thermocouples directly affixed to the crystal base in each monochromator. 
The power absorbed by the first Si crystal $P$ was calculated using mass-energy x-ray absorption coefficient of Si
and equations for synchrotron radiation \cite{xdb2001,hbook_srad1A} with the relevant parameters of a bending magnet and undulator A at the Advanced Photon Source. The main experimental parameters are listed in Table~\ref{tab:oc} for each experiment. 

\begin{table} 
\label{tab:oc}
\caption{Experimental parameters during x-ray diffraction imaging experiments of double-crystal high-heat-load monochromators at 1-BM and 20-ID beamlines of the Advanced Photon Source: \\
$L$ - distance from the source to the monochromator, \\
$D$ - distance from the monochromator to the observation plane, \\
$E_0^{111}$ - photon energy of the working Si 111 reflection, \\
$P$ - total radiation power absorbed in the Si crystal, \\
$\delta y$ - spatial resolution in the tangential direction (Eq.~\ref{eq:dy}), \\
$T$ - measured reference temperature on the first monochromator crystal.
} 
\begin{tabular}{l c c c c c c} 
Beamline  &  $L$ [m]  &  $D$ [m] & $E_0^{111}$ [keV] & P [W] &  $\delta y$ [$\mu$m] & T [K]\\
1-BM      &  27.5     &  7.0     & 6.0               & 1.3      &   58              & 289  \\
20-ID     &  30.0     &  20.0    & 11.0              & 93.5     &   84              & 82   \\
\end{tabular} 
\end{table} 

\section{Analysis and Discussion}
\begin{figure*}
\includegraphics[scale=0.4]{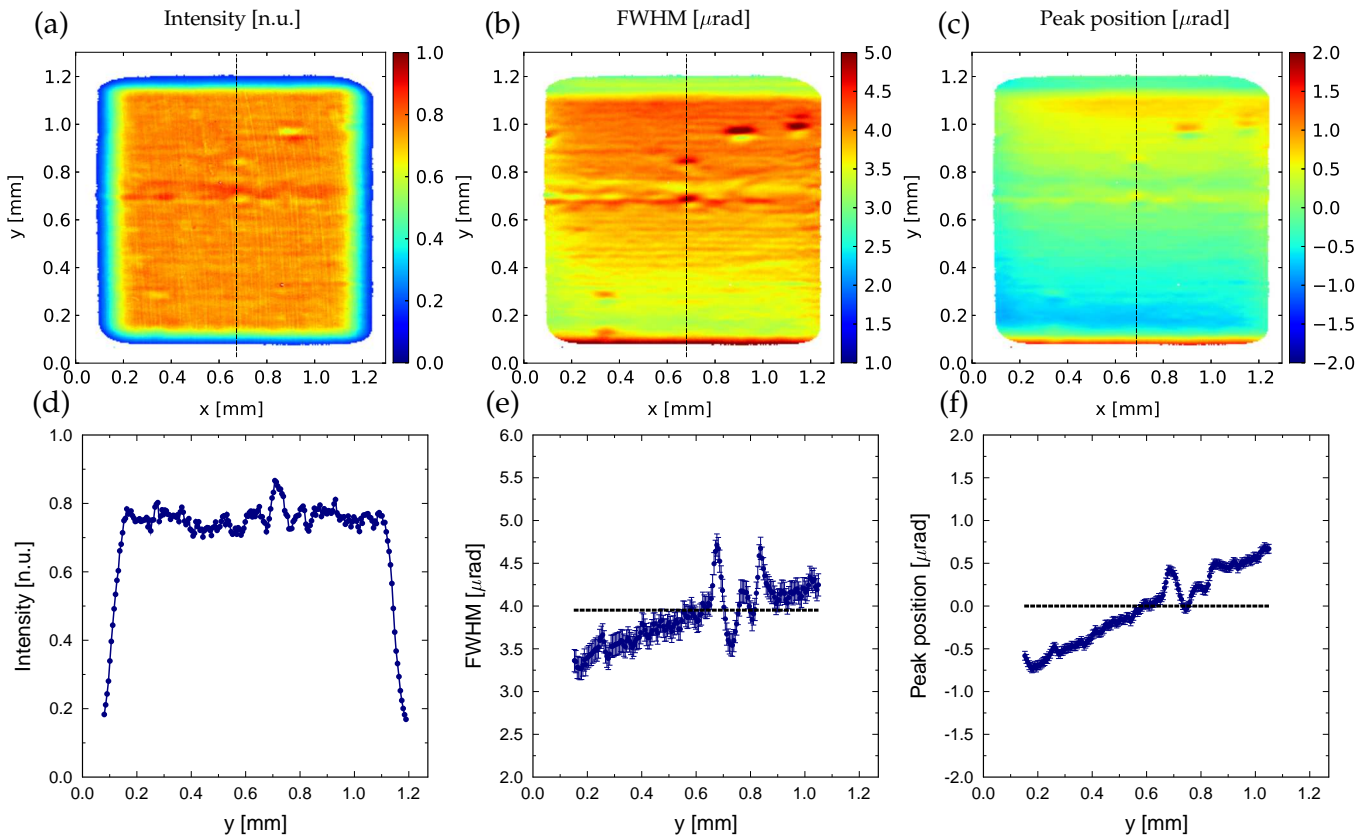} 
\caption{Rocking curve topographs of the exit beam of a double-crystal monochromator (tuned to a photon energy of $E_0^{111} = 6$~keV) at a bending magnet beamline of the Advanced Photon Source (1-BM): 
a map of the normalized peak intensity of the local rocking curves (a), a map of the curve width (b) and a map of the peak position (c). 
Each map is sliced as shown by the vertical dashed line. The resulting corresponding distributions along these lines are those of peak intensity (d), curve width (e) and peak position (f). The theoretical value of the rocking curve width for perfect crystals in the double-crystal geometry (4~$\mu$rad) is shown by the dashed line in (e). The dashed line in (f) represents the reference zero level. The size of the incident x-ray beam is limited by 1x1 mm$^2$ aperture upstream of the monochromator.} 
\label{fig:1bm}
\end{figure*} 

\begin{figure*}
\includegraphics[scale=0.4]{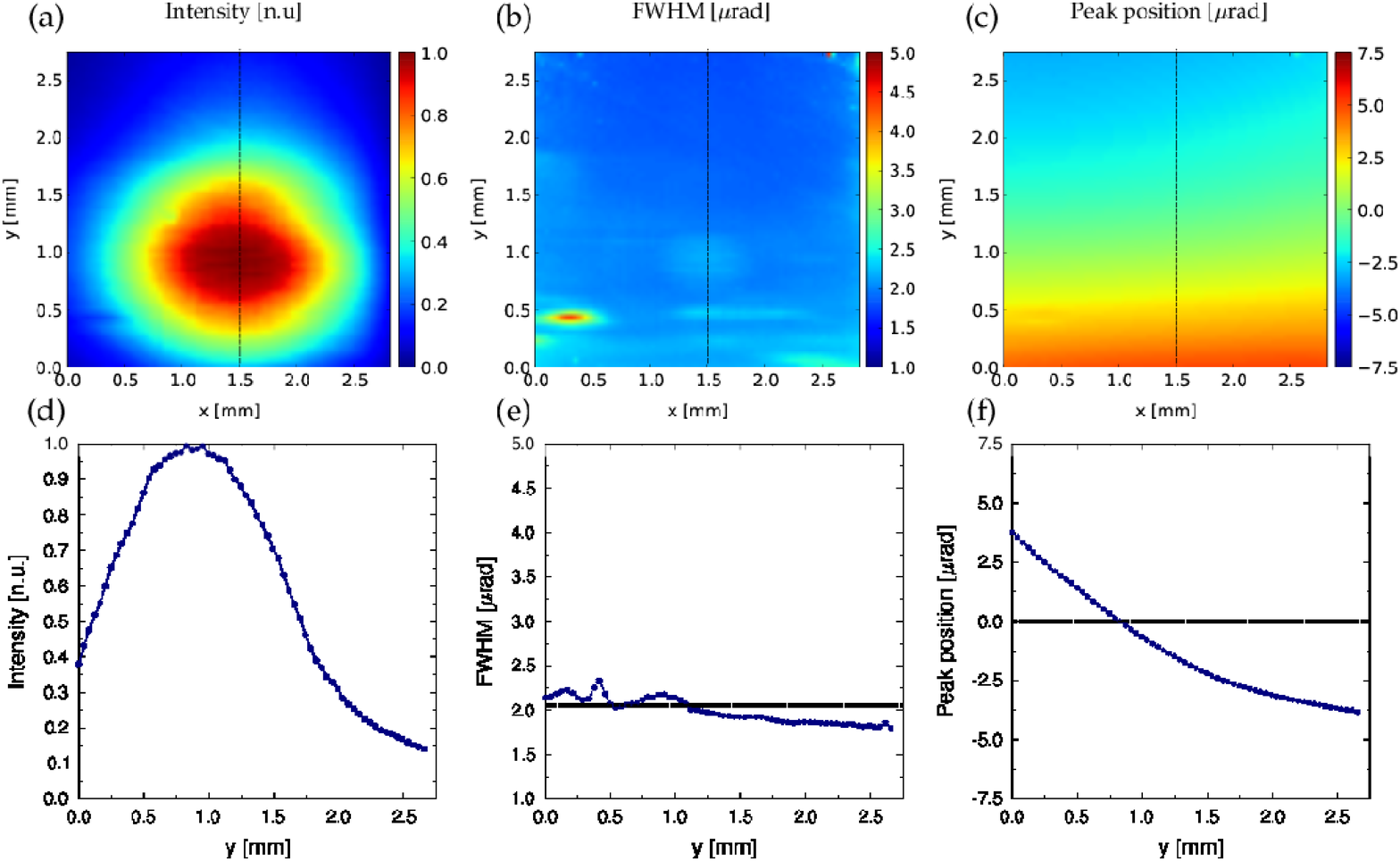} 
\caption{Rocking curve topographs of the exit beam of a cryo-cooled double-crystal monochromator at an undulator beamline of the Advanced Photon Source (20-ID) for the case of moderate heat load ($E_0^{111}$ = 11~keV, undulator deflection parameter $K$ = 0.75, total absorbed power $P$ = 93.5 W):
a map of the normalized peak intensity of the local rocking curves (a), a map of the curve width (b) and a map of the peak position (c). 
Each map is sliced as shown by the vertical dashed line. The resulting corresponding distributions along these lines are those of peak intensity (d), curve width (e) and peak position (f). The theoretical value of the rocking curve width for perfect crystals in the double-crystal geometry (2.2~$\mu$rad) is shown by the dashed line in (e). The dashed line in (f) represents the reference zero level. A portion of the observation plane is shown which includes the central radiation cone. The total observation area is limited by an entrance aperture of 2.4$\times$1.2~mm$^2$ upstream of the monochromator.} 
\label{fig:20id}
\end{figure*}

The collected images were sorted to calculate local rocking curves for each pixel. Sequential x-ray topographs were computed using rctopo code of DTXRD package~\cite{dtxrd}. 
To reduce noise in the topographs the parameters of rocking curves at each pixel were extracted using fitting with the Gaussian function. 
Figures~\ref{fig:1bm}(a-c) show sequential x-ray topographs of the first experiment with the water-cooled double-crystal monochromator at the bending magnet beamline. Each topograph is sliced as shown by the vertical dashed line. The resulting distributions are plotted in Figs.~\ref{fig:1bm}(d-f) accordingly. 
Figures~\ref{fig:1bm}(a,d) show distributions of the normalized peak intensity of local rocking curves. Aside from a few local disturbances the peak intensity is quite uniform as expected for high-quality Si monochromator crystals at low heat load.  
Distributions of the rocking curve width shown in Fig.~\ref{fig:1bm}(b,e) reveals more detailed information on the local disturbances suggesting presence of intrinsic localized strain in several locations (increased local FWHM as compared to the theoretical value of 
$\Delta \theta_2 \simeq$~4~$\mu$rad for Si 333 double-crystal rocking curve at 18~keV). 
For small regions of interest such as the one defined by the entrance aperture these disturbances can be avoided using crystal translations.  
Distributions of the rocking curve peak position shown in Fig.~\ref{fig:1bm}(c,f) reveal a small gradient of $\simeq 1 \mu$rad peak-to-valley. We note that in the case of water-cooled monochromator crystal the thermal expansion coefficient of Si near the room temperature is quite substantial $\alpha \simeq 2.6 \times 10^{-6}$. Small temperature variations along the beam footprint (e.g., $\delta T \approx$~1~K) can lead to a substantial variation in the d-spacing $\delta d_H/d_H = \alpha \delta T \approx 10^{-6}$. Therefore, the angular deviation defined by Eq.~\ref{eq:dth_dd} can no longer be neglected. Thus, at room temperatures the measured gradient can not be solely attributed to the misorientation of the Bragg planes. Nevertheless, regardless of the origin the measured gradient represents the disturbance in the wavefront due to non-ideal state of the crystal. 


Figures~\ref{fig:20id}(a-c) show sequential x-ray topographs of the second experiment with the cryo-cooled double-crystal monochromator at the undulator beamline. Similarly to the previous case each topograph is sliced as shown by the vertical dashed line. The resulting distributions are plotted in Figs.~\ref{fig:20id}(d-f) accordingly. 
The topographs show a portion of the imaging plane which includes the central radiation cone. 
The map of the rocking curve peak intensity Figures~\ref{fig:20id}(a) represents the envelope of the Si 333 reflected intensity. At each particular angle on the rocking curve the reflected intensity is present only in a narrow stripe in the imaging plane as shown by the sequence of alternating images (Supplementary section). This stripe moves in the vertical direction as angle of the second crystal is scanned. This behavior manifests itself as a strong gradient shown in Figs.~\ref{fig:20id}(c,f) ($\approx$~8~$\mu$rad peak-to-valley). The width of the local rocking curves is quite uniform (Figures ~\ref{fig:20id}(b,e)) and close to the theoretical value $\Delta \theta_2$~=~2.2~$\mu$rad. This indicates high crystal quality over the entire observation region.

\begin{figure*}
\includegraphics[scale=0.4]{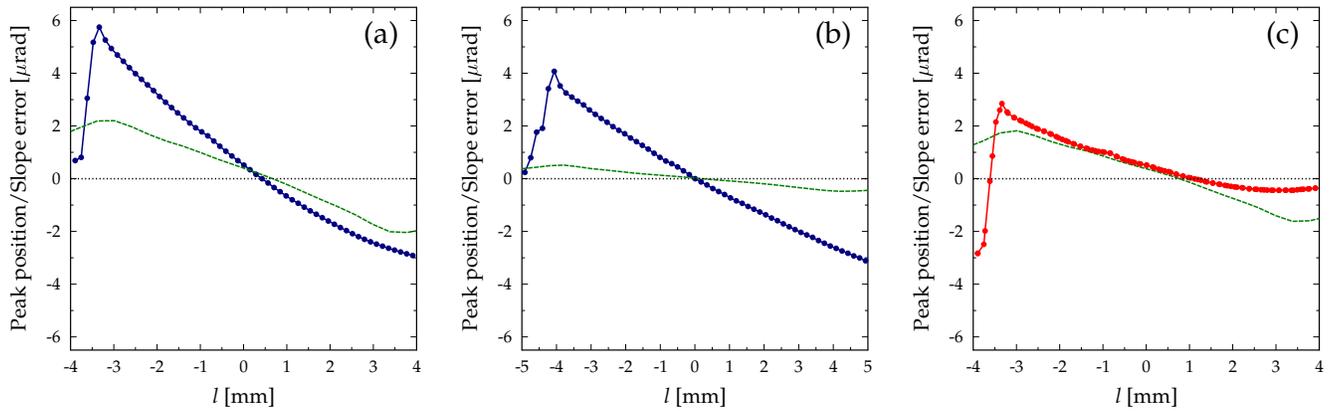} 
\caption{Angular profiles along the center line of the beam footprint on the first crystal. The results of finite element analysis are shown by green dashed line and the measured angular profiles are shown by circles and solid lines. The angular profiles represent the concave shape of the crystal. The peak in the measured profiles is due to clipping of the reflected intensity by the bottom edge of the entrance aperture. The peak was used for alignment purposes.
(a) the first undulator harmonic is tuned to a photon energy of 11~keV; (b) the first undulator harmonic is tuned to a photon energy of 13~keV. 
(c) the experimental angular profile of (b) is taken as a reference level and subtracted from the experimental profile of the case (a). The resulting difference curve (red circles, red solid line) is compared with the difference between the corresponding results of FEA (green dashed line). 
} 
\label{fig:fea}
\end{figure*}

As discussed in section~\ref{sec:theory} the measured angular profile of the rocking curve peak position should represent the distribution of heat-load-induced misorientation of the Bragg planes or the surface slope error given that the initial state of the crystal in the absence of heat load is strain-free. 
It is imperative to compare the measured slope error with predictions of finite element analysis (FEA). 
A volumetric heat source in the Si crystal produced by the undulator beam was obtained using equations describing spectral/spatial distribution of undulator radiation~\cite{hbook_srad1A} and the energy-dependent x-ray absorption coefficient of Si. The heat source was introduced in finite element analysis with boundary conditions consistent with the typical scheme of an indirect-cooled Si monochromator~\cite{ZLiu14}.  
The cooling channel of the monochromator crystal holder had wire-coil inserts for enhanced heat transfer such that the effective transfer coefficient was assumed to be 8~kW/m$^2$K$^{-1}$ \cite{Collins02}. The heat transfer in the silicon-copper interface with indium foil as the interstitial material is known to depend on the pressure applied~\cite{Khounsary97}. The effective value of 4~W/m$^2$K$^{-1}$ was assumed for this parameter in the performed finite element analysis. The slope error along the center line of the beam footprint on the first crystal extracted from the results of FEA is shown in Fig.~\ref{fig:fea} by the green dashed lines. 

The measured angular profile of the central slice (Figure~\ref{fig:20id}(f)) was recalculated in the coordinate along the surface of the crystal ($l$). The profile was aligned with respect to the reference zero level of the central radiation cone using the bottom edge of the aperture, which was clipping the x-ray beam incident on the monochromator. The angular profile is shown in Fig.~\ref{fig:fea}(a) by circles and a solid line. 
We note that the positive sense of rotation $\delta \theta$ defined as shown in Fig.~\ref{fig:setup} resulted in the negative (top-to-bottom) shift of the exit beam in the observation plane (see Supplementary material). The exit beam shifted to the bottommost position represents the upstream portion of the first crystal. According to the diagrams shown in Fig.~\ref{fig:rtrace}(c) for a convex shape of the crystal surface the direction of the beam shift should be opposite. Thus, the actual shape of the crystal in the experiment was concave. The resulting angular profile is plotted in Figure~\ref{fig:20id}(f). It represents the derivative of the crystal surface profile (i.e., the slope error). The surface profile in our analysis is considered with respect to the inward normal to the crystal surface. 

The result of FEA correctly predicts the shape yet underestimates the measured profile. In an attempt to analyze the origin of this discrepancy we established that the temperature of the base of the first crystal during the experiment was $\approx$~82~K which agreed with the result of FEA. The maximum increase in the temperature with respect to this base level predicted by FEA was 10~K. It was concluded that the discrepancy between the angular profiles can not be attributed to the resulting change in the d-spacing ($< 1 \mu$rad) induced by such small temperature deviation at this low temperature. 
The discrepancy could be attributed to a residual strain due to mounting of the crystal or thermal destabilization of strain-free mounting conditions due to differential thermal expansion upon cooling to the cryogenic temperatures. Figure~\ref{fig:fea}(b) shows the angular profile measured at low heat load (open gap of the undulator corresponding to a photon energy of 13~keV) compared to FEA slope error profile calculated for this condition (absorbed power 24~W). The result of FEA suggests that the heat-load-induced slope error can be neglected. Thus, the measured angular profile represents a reference level due to residual mounting-induced slope error which can be subtracted from the result of measurement at the moderate heat load (Figure~\ref{fig:fea}(a)) to yield the effective heat-load-induced slope error. This difference shown in Figure~\ref{fig:fea}(c) by circles and the solid line is compared with the
difference between the corresponding FEA-calculated profiles (dashed green line). As seen from the figure such renormalization procedure yields a good agreement except at the downstream most portion of the beam footprint where an asymmetric deviation from the FEA slope error is observed. This asymmetry could be attributed to imperfect alignment of the entrance aperture. Such a likely conclusion is supported by the fact that only the bottom beam-clipping edge of the aperture was observed in the experiment.

\section{Conclusions}
In conclusion, x-ray diffraction imaging method was applied for rapid in-situ diagnostics of double-crystal monochromators. The method is based on sequential x-ray diffraction topography of high-order Bragg reflections present in the exit beam of double-crystal monochromators at synchrotron radiation facilities. The method provides visualization of the working crystal region and its properties under the heat load of intense synchrotron radiation. The state of the thermally distorted first crystal is monitored using local rocking curves with a spatial resolution which can be as low as $\approx$~10~$\mu$m. Parameters of local rocking curves are mapped across the observation region and can be interpreted as a map of the beam footprint (reflected intensity), local crystal quality (curve width) and wavefront distortion (peak position). In the case of cryo-cooled working crystals the latter can be attributed to local misorientation of the reflecting crystal planes which can be compared to heat-load-induced slope errors derived from finite element analysis. The characterization procedure at any given heat load condition involves a single angular scan of the second monochromator crystal and data analysis using a software code. The procedure can be easily automated. Rapid complete evaluation of the double-crystal monochromators provided by the method can facilitate development of adaptive systems where the state of the crystal is manipulated to minimize heat-load-induced wavefront distortion of the exit beam. 
\begin{acknowledgments}
We thank M. Beno, G. Navrotski, J. Lang, A. Macrander and other members of the high-heat-load monochromator working group at the Advanced Photon Source for helpful discussions. L. Berman is acknowledged for valuable comments.
M. Moore, R. Woods and M. Pape are acknowledged for technical support.  
Use of the Advanced Photon Source was supported by the U. S. Department of Energy, 
Office of Science, Office of Basic Energy Sciences, under Contract No. DE-AC02-06CH11357.
\end{acknowledgments}
\newpage


%
\end{document}